\documentclass[twocolumn,superscriptaddress,showpacs,10pt,prl]{revtex4}

\usepackage{graphicx}%
\usepackage{dcolumn}
\usepackage{amsmath}
\usepackage{ifthen}

\newboolean{camera}

\setboolean{camera}{true}\makeatletter\appdef\set@pica@hook{\textheight = 22.5cm}\makeatother

\DeclareGraphicsExtensions{.eps}
\graphicspath{{bilder/}}

\ifthenelse{\boolean{camera}}{
\setlength{\textwidth}{17cm}
\setlength{\columnsep}{1cm}
\addtolength{\topmargin}{-.5cm}
\addtolength{\headsep}{1.5cm}
\makeatletter
\def\frontmatter@abstractwidth{\textwidth}
\makeatother}{}

\makeatletter
\def\@pacs@name{Keywords: }
\makeatother

\begin{document}

\ifthenelse{\boolean{camera}}{
  \title{\vspace*{1cm}
  Phonon emission and absorption in the fractional quantum Hall effect
 \vspace*{-1em}}}
{\title{Phonon emission and absorption in the fractional quantum Hall effect}}

\author{F.~Schulze-Wischeler}
\author{U.~Zeitler}
\email[corresponding author, e-mail: ]{zeitler@nano.uni-hannover.de}
\author{M.~Monka}
\author{F.~Hohls}
\author{R.~J.~Haug}
\affiliation{%
Institut f\"ur Festk\"orperphysik,
Universit\"at Hannover, Appelstra{\ss}e 2, 30167 Hannover, Germany
}%

\author{K.~Eberl}
\affiliation{
MPI f\"ur Festk\"orperforschung, Heisenbergstr.~1, 
70569 Stuttgart, Germany
}%

\ifthenelse{\boolean{camera}}{}{\date{\today}}

\newcommand{\mykeywords} {fractional quantum Hall effect, phonons, energy
relaxation, magneto-roton gap}

\begin{abstract}

We investigate the time dependent thermal relaxation of a two-dimensional
electron system in the fractional quantum Hall regime where ballistic 
phonons are used to heat up the system to a non-equilibrium temperature. 
The thermal relaxation of a 2DES at $\nu=1/2$ 
can be described in terms of a broad band emission of phonons, with 
a temperature dependence proportional to $T^4$. 
In contrast, the relaxation at 
fractional filling $\nu=2/3$ is characterized by phonon emission around
a single energy, the magneto-roton gap. This leads to a strongly
reduced energy relaxation rate compared to $\nu=1/2$ with only a 
weak temperature dependence for temperatures 150~mK $< T <$ 400 mK.

\ifthenelse{\boolean{camera}}{}
{ \vspace{1em} Keywords: \mykeywords\\[1em]
  \noindent
   Corresponding author: \parbox[t]{12cm}{Dr.\ U.\ Zeitler,\\
        Institut f\"ur Festk\"orperphysik, Universit\"at Hannover\\
        Appelstra{\ss}e 2, 30167 Hannover, Germany\\
        fax: +49 (0)511 768-2904\\
        tel: +49 (0)511 768-2903\\
        e-mail: zeitler@nano.fkp.uni-hannover.de}
}

\end{abstract}

\pacs{\mykeywords}

\maketitle

\newlength{\plotwidth}
\setlength{\plotwidth}{0.7\linewidth} 
\ifthenelse{\boolean{camera}}{\thispagestyle{myheadings}
   \setlength{\plotwidth}{\linewidth}}{} 

It has been shown recently that phonon absorption experiments
are an efficient tool for the investigation of the thermodynamic
and spectroscopic properties of highly correlated 
two-dimensional electron systems (2DESs) in the fractional quantum Hall (FQH) 
regime~\cite{uliPRL}. In this work we will present first
results on the phonon emission of a 2DES. We use ballistic phonons to
heat up the 2DES to a non-equilibrium temperature and study the 
thermal relaxation back to equilibrium. 
At filling factor $\nu = 1/2$ we find a fast energy relaxation
rate proportional to $T_{ne}^4 - T_0^4$, where $T_{ne}$ is the 
non-equilibrium temperature of the 2DES and $T_0$ is the
equilibrium temperature of the substrate. For fractional filling
$\nu=2/3$ the energy relaxation rate is much less temperature
dependent indicating the dominance of magneto-roton relaxations in
the thermal relaxation process.

Our samples were grown by molecular beam epitaxy (MBE) on a 0.53~mm thick GaAs 
wafer. Both sides were polished to an optical finish ready for epitaxy
before the growth. 
We have performed phonon experiments on the 2DES of two different samples.
Sample 1 has a electron concentration $n = 1.1 \times 10^{11}$~cm$^{-2}$,
and a mobility $\mu = 6 \times 10^5$~cm$^2$/Vs, for sample 2 we have
$n = 1.3 \times 10^{11}$~cm$^{-2}$ and $\mu = 6 \times 10^5$~cm$^2$/Vs.
The 2DESs are patterned in form of a meander on a $1 \times 1$-mm$^2$ square.
To increase sensitivity to changes in the resistivity $\rho_{xx}$ the 
meander has a large length-to-width ratio, $L/w =2000$ for sample 1
and $L/w=360$ for sample 2.

The samples were mounted in vacuum on the tail of a dilution-refrigerator.
They are thermally connected to the mixing chamber by a rod made of Cu-wires.
The two contacts of the 2DES meander are connected to 
a 50-$\Omega$ coaxial line. To avoid heat leaks through the
center conductors of the coaxial lines they are
thermally anchored to an impedance-matched strip-line on the mixing-chamber
before connecting it to the sample.
One contact of the sample is biased with a DC-voltage of
a few mV. The other contact is connected via a 5-M$\Omega$ resistor close
to the sample to ground. Additionally it is AC coupled to a
50-$\Omega$ current amplifier
at the room temperature end of the coaxial line.
Under DC conditions a constant current of typically 50 nA passes through 
the sample and the 5-M$\Omega$ bias resistor.
The current is small enough
to minimize Joule heating of the 2DES but still large enough to
get a reasonable phonon signal (see below).

On the rear face of the GaAs substrate a constantan heater was evaporated.
The heater is impedance 
matched to a 50-$\Omega$ coaxial line which connects it to the pulse
generator. Again the line is thermally anchored
to an impedance-matched strip-line before connecting it to the heater.

Non-equilibrium phonons are created by passing an electric pulse through the
constantan heater. They traverse the substrate ballistically and
hit the 2DES. A small part of the phonons are absorbed which
leads to an increase of the 2DES temperature. As a consequence,
its resistance changes and a time dependent transient current is
created. This current is detected by the AC-coupled current amplifier
with a 10~MHz bandwidth limiting the time resolution of our setup 
to about 40~ns. 

The resistance change as function of time is recorded with a
digital storage oscilloscope connected to the output of the
current amplifier. To increase the signal to noise ratio the curves
are averaged over typically $10^4$ traces with a pulse repetition 
period of 10-100 ms. The repetition period has to be that low to allow
the whole sample holder to relax to the base temperature of
the mixing chamber before a further phonon pulse is applied
and to minimize the average heating effects on the sample holder.

\begin{figure}[t] 
  \begin{center}
  \resizebox{0.9\linewidth}{!}{\rotatebox{270}{\includegraphics{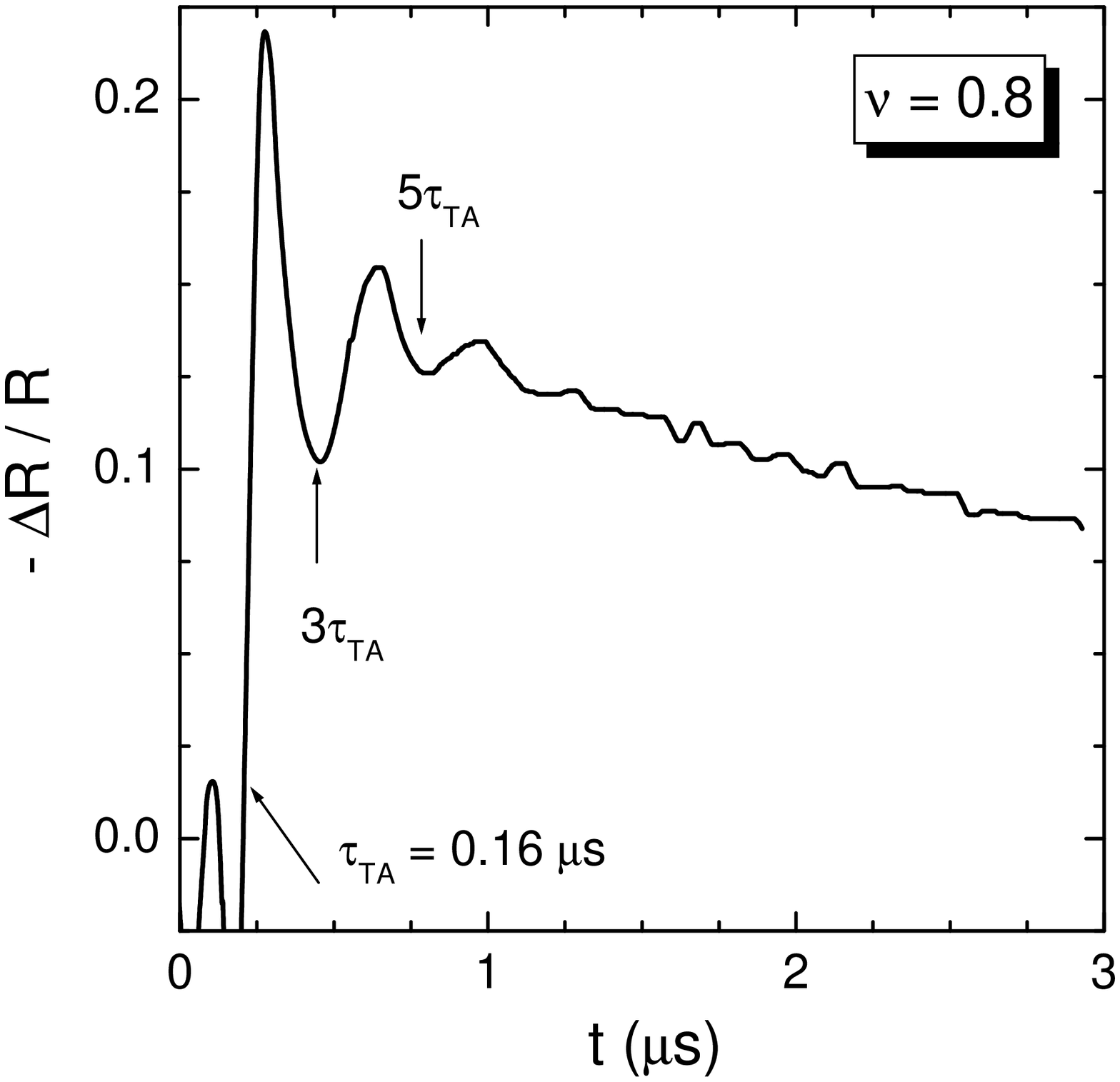}}}
  \end{center}
  \caption{Relative change of the 2DES resistance as a function of time
           at $\nu = 0.8$ after emission of a 50-ns phonon pulse on the 
           rear face at $t=0$. 
           The arrows mark the expected
           transient times of ballistic TA-phonons through the 0.53-mm thick
           substrate. Note that the relative resistance changes are negative.}
  \label{fig1}
  \vspace*{1em}
\end{figure}

A typical phonon signal for sample 1 is shown in Fig.~1. The measurements 
were performed at a base temperature of the mixing chamber $T_{mc} = 50$~mK.
At the chosen filling factor $\nu=0.8$ the 2DES resistance is extremely
sensitive on temperature. As a consequence small changes in $T$ show
up as large changes in the resistance as shown in Fig.~1. 
A 50-ns phonon pulse is emitted at $t=0$. The observed signal 
at small times is an artifact originating from electromagnetic pick-up
of the heater pulse by  the detection system (2DES and contacting wires).
This effects of pick-up are already
minimized by alternating the pulse amplitude and the direction of
the bias current. However, due to the thinner substrate compared
to previous work~\cite{uliPRL,AndyEP2DS} they still can lead
to a significant unwanted contribution to the phonon-signal.
Therefore, at filling factors where the 2DES is less sensitive
to phonons, the pick-up can even mask the ballistic contribution
originating from the clean phonon signal. 

After $\tau_{TA} \approx 0.16~\mu$s transverse acoustic phonons have 
traversed the substrate and hit the 2DES. Longitudinal
acoustic phonons only play a minor role in the phonon-absorption
process.
As a consequence of the phonon absorption
the 2DES resistance $R$ decreases drastically and starts to relax
back to its original value as soon as no more ballistic phonons
are incident on the 2DES. Further decreases of $R$ occur when the
ballistic phonons which were reflected on the front and rear side
of the sample hit the 2DES again after $3\tau_{TA}$ and $5\tau_{TA}$.
For longer times ($t>1 \mu$s) most of the ballistic phonons
have thermalized and the ballistically heated 2DES relaxes back
to the substrate temperature. 
At long time scales ($t \approx 10~\mu$s) the energy dissipated in the 
heater is equally distributed in the whole GaAs substrate 
leading to a higher substrate temperature 
compared its value before the phonons were emitted. 
It takes a few 100~$\mu$s for the equilibrium phonons
responsible for this background heating to leave the GaAs substrate.

The effects of the ballistic heating of the 2DES have already 
been described in previous work~\cite{uliPRL,AndyEP2DS}. In this
paper we concentrate on the relaxation of the ballistically heated 2DES.
In this respect the precise phonon absorption mechanism does not play
an important role for the understanding of our results. The phonons
are simply used to create a hot 2DES on very fast time scales of
a few ns with a comparably negligible heating of the substrate. 
By this we can directly study the time dependent thermal relaxation
of a rapidly heated 2DES back to the substrate temperatures.

\begin{figure}[t]
  \begin{center}
  \resizebox{0.9\linewidth}{!}{\rotatebox{270}{\includegraphics{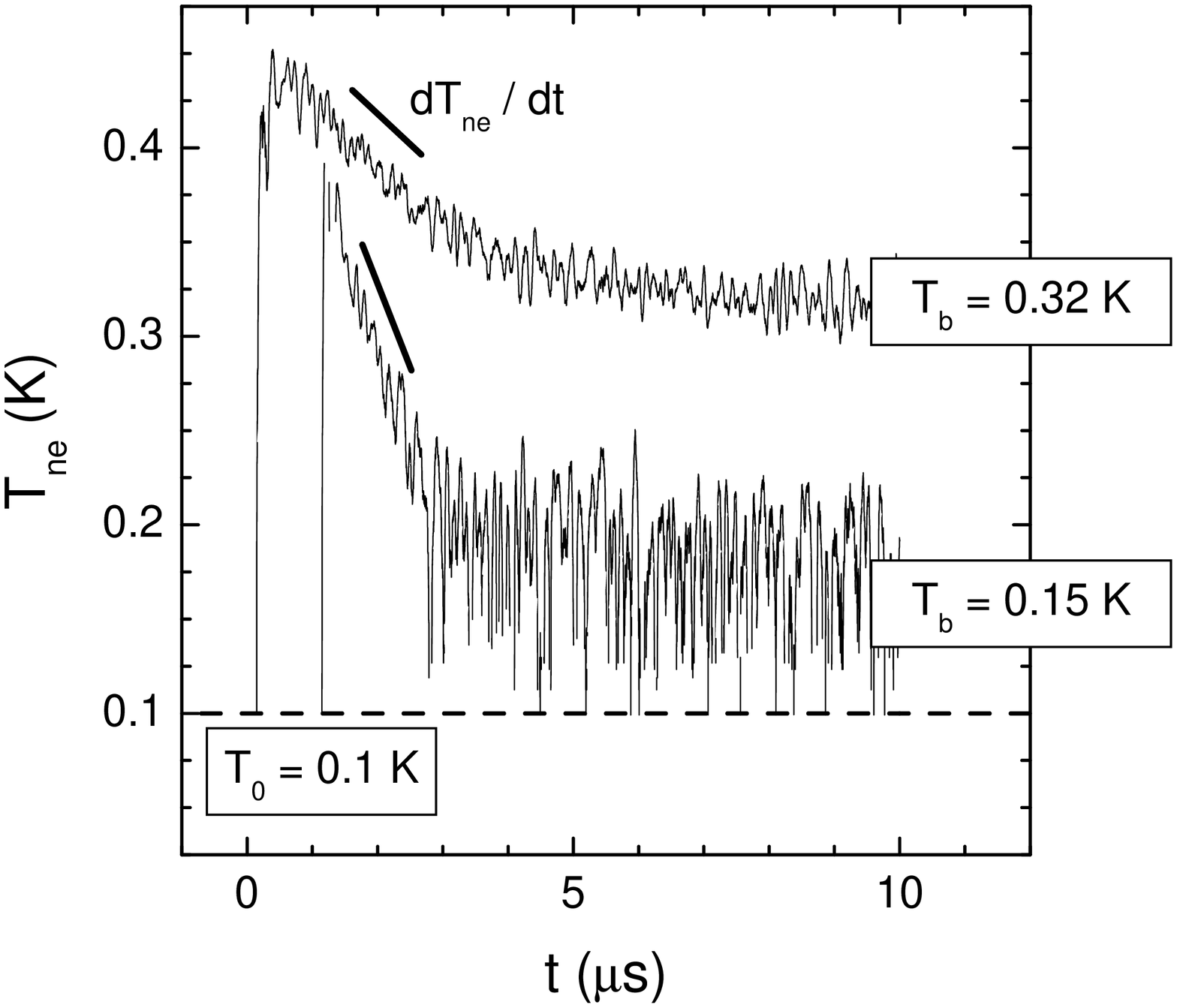}}}
  \end{center}
  \caption{Time dependent relaxation of the non-equilibrium 
          2DES temperature at $\nu=2/3$ after heating the system 
          by a ballistic phonon pulse characterized by a non-equilibrium
          phonon temperature $T_h = 2$~K (top trace) and $T_h = 0.9$~K 
          (bottom trace). For clarity the bottom-trace is shifted to  
          the right by 1~$\mu$s. Artifacts originating from
          electromagnetic pick-up have been removed.
          After the time of flight for TA-phonons, the 2DES rapidly
          warms up and relaxes slowly to the substrate temperature $T_b$.}
  \label{fig2}
\vspace*{1em}
\end{figure}

Such a relaxation experiment is shown in Fig.~2 for the fractional
filling factor $\nu =2/3$ (sample 2) for two different
heater temperatures $T_h$. The 2DES is initially
at a temperature $T_0 = 100$~mK. At $t=0$ a 50-ns phonon pulse
characterized by a non-equilibrium temperature $T_h$ is emitted
from the heater at the back face of the substrate. 
After hitting the 2DES, the electron
system is heated up rapidly by the ballistic phonons and then 
slowly relaxes back. The non-equilibrium temperature $T_{ne}$ of
the electron system is calculated from a calibration of the sample 
resistance $R(T)$ under equilibrium conditions.
As can been seen more clearly in Fig.~1 it
is secure to assume that the  time dependence of $T_{ne}$ is mainly
determined by phonon emission, i.~e.~absorption processes due multiple
reflections have sufficiently decayed.

After a few microseconds the 2DES has relaxed to the substrate
temperature $T_b$ defined by the total specific heat of 
the GaAs substrate and the heater temperature $T_h$ 
(see \cite{uliPRL} for more details).

To get more information about the relaxation process  
we have evaluated the change of the non-equilibrium temperature
as a function of time, $dT_{ne}/dt$, at $t=1.5~\mu$s where the 2DES 
is at a temperature $T_1$. 
At this time all possible artifacts in the phonon-signal
caused by electromagnetic pick-up
have sufficiently decayed.
The results for filling factors
$\nu=1/2$ and $\nu=2/3$ are shown in Fig.~3 where $dT_{ne}/dt$ 
measured at $t=1.5~\mu$s is plotted as a function of $T_1$.. 
 
At filling factor $\nu = 1/2$ the relaxation rate increases 
drastically with increasing $T_1$ as indicated with the dashed line
in Fig.~3a. The data available roughly follow a $T^3$ dependence.

\begin{figure}[t]
  \begin{center}
  \resizebox{0.99\linewidth}{!}{\includegraphics{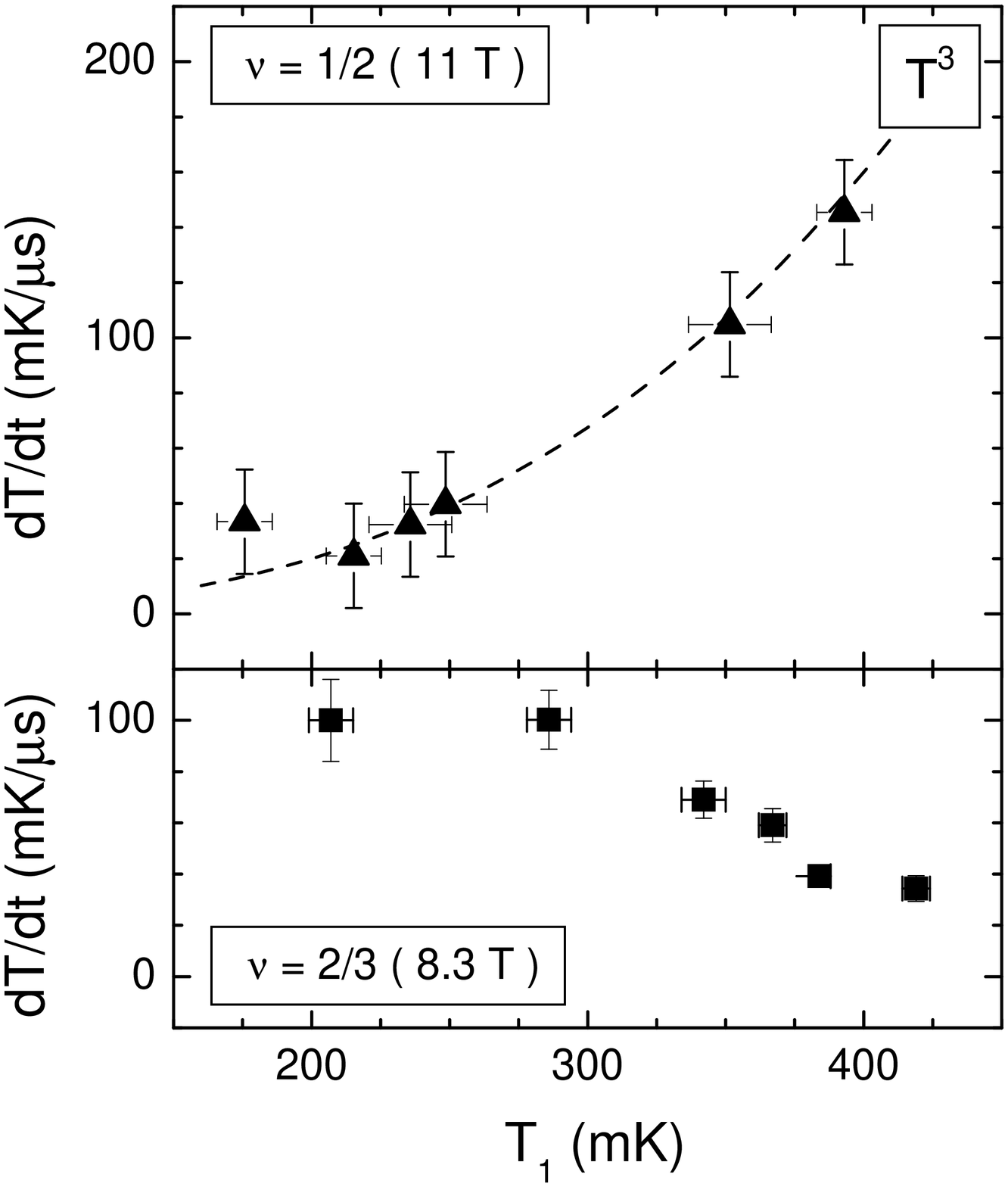}}
  \end{center}
  \caption{Thermal relaxation of a 2DES at filling factor 
   $\nu=1/2$ (top panel) and $\nu=2/3$ (bottom panel) as described
   in the text.}
  \label{fig3}
\vspace*{1em}
\end{figure}

For a more quantitative analysis it is more convenient to regard
the energy relaxation rate, $dE/dt$, rather then the change $dT_{ne}/dt$.
The energy change $dE$ is related to $dT_{ne}$ as
$dE= C(T) dT_{ne}$. For $\nu=1/2$ is 
reasonably to assume specific heat is linear in T, and, as a consequence
$dE/dt$ is roughly proportional to $T_{ne}^4$. This is the expected 
behavior for the energy relaxation of a simple metallic film emitting
a black-body spectrum of phonons. 
The experimental accuracy at present does not allow to 
determine the exponent in the power law accurately enough and to
compare it to the temperature dependence of phonon interaction with 
electrons in the fractional quantum Hall regime as deduced
for transport experiments~\cite{Kang}, thermopower~\cite{Benno}
and phonon emission~\cite{Chow}. 

In contrast to $\nu = 1/2$ the relaxation rate $dT_{ne}/dt$
at $\nu = 2/3$ {\sl decreases} with increasing T.
In other words the hotter the 2DES gets the slower it relaxes~\cite{rem2}.
To explain the origin of this peculiar behavior we propose two possible 
mechanisms for the effect observed.

First, the energy relaxation rate might follow a much weaker power law 
than $T^4$.
E.g.~an energy relaxation relaxation rate proportional to $T$ with an
electronic specific heat $C_V \propto T$ would yield a $dT_{ne}/dt$
which is independent of temperature. Therefore, any temperature
dependence of $dE/dT$  weaker than linear causes a $dT_{ne}/dt$
which is decreasing with increasing $T$.

Second, the specific heat $C(T)$ of a 2DES at fractional
filling factors is only linear in $T$ for very low temperatures.
In Ref.~\cite{uliPRL} strong deviation from a linear 
behavior were found for $T>T^* \approx 0.4~$K at $\nu=1/3$. 
The temperature $T^*$ is related to the width of the $\nu=1/3$
minimum in the density of states. For our case where we investigate
the $\nu=2/3$ minimum (at a lower magnetic field in a sample with
a smaller mobility) a lower $T^*$ than 400~mK is expected. 
The strong decrease of $dT_{ne}/dt$ as a function of $T$ is  
therefore an indication for a strongly increasing $C(T)$.

The weak temperature dependence of $dE/dT$ points to the 
fact that the emission process at fractional filling
is distinct from the weakly interacting composite-fermion 
system at $\nu=1/2$. It is indeed to expect that the emission
process is dominated by phonon emission at a single energy, namely the
magneto-roton gap. 

In conclusion we have investigated the phonon emission of a hot
2DES in the FQH regime using time-resolved resistance measurements.
We find first indications that the emission process is
governed by the radiation of a black body spectrum at $\nu=1/2$
whereas at $\nu=2/3$ an emission of phonons around a single 
energy is suggested by our experiments.
   
This work is supported by the Deutsche Forschungsgemeinschaf,
project no.~ZE 463/3-1.



\begin{thebibliography}{99}
\bibitem{relax}
Strictly speaking the relaxation rate is of course given
by $dE/dT \propto (T_{ne}^4 - T_b^4)$. However, within our accuracy
this modification does not change our qualitative arguments.
\bibitem{rem2}
Again the energy relaxation rate is strictly spoken proportional
to $T_{ne}^\alpha - T_b^\alpha$ with a unknown exponent $\alpha$.
For high $T_b$ and low $\alpha$ the second term can no more
be neglected. However, it should be mentioned that we only use
qualitative arguments to describe the effects observed. Including
the above modification complicates the situation but does not
give additional information.
\end{thebibliography}
\end{document}